# More satellites around the nearby spiral galaxy NGC 7331

M. Blauensteiner[1], I. D. Karachentsev[2], P. Remmel[1], P. Riepe[1], M. E. Sharina[2], H. Strauß[1], U. Trulson[1], T. Zilch[1]

**Abstract**
We report the detection of two dwarf galaxies in a projected distance of ~50 kpc from NGC 7331 and suspect the physical nature of dwarfs of this spiral galaxy.

The far-off outskirts of massive galaxies contain traces of ancient events that took place thousands of million years ago. Investigations of our neighbour galaxy M 31´s periphery revealed several stellar streams, remnants of satellite dwarf galaxies destroyed by tidal forces [1, 2]. The brightness of these streams is only a small fraction of the moonless night sky´s brightness. According to the standard cosmological model there should be numerous smaller galaxies around the giant galaxy, their expected numbers should be much greater than presently observed [3]. It is accepted that up till today many of these satellite galaxies could not be found because of their very low surface brightness. In recent years the search for such "lost satellites" has got more importance. There have been established various groups of professional and amateur astronomers, successfully hunting for low surface brightness objects and structures in the environment of nearby massive galaxies [4-8]. In order to reach such faint targets, large telescopes as well as commercial small ones are used in the optical spectral range, operating with very long exposure times.

In this message we present results from our searching for faintest structures in the surrounding of the massive spiral galaxy NGC 7331. This galaxy resembles a twin of M 31, compared by luminosity, extension, rotational amplitude and morphology. The cepheid based distance to NGC 7331 is 14.5 Mpc [9], the radial velocity $V_{LG}$ relatively to the centre of the Local Group 1119 km/s. In agreement with [10] NGC 7331 is the main member of a group of four galaxies with a radial velocity dispersion of just 44 km/s, a virial radius of 276 kpc, a summed up stellar mass of $M^* = 2.0 \times 10^{11}$ $M_\odot$ and a virial mass of $2.2 \times 10^{12}$ $M_\odot$. Recently, Ludwig et al. [11] investigated the surrounding field of NGC 7331 with the 0.9-m telescope of Kitt Peak National Observatory with a total exposure of 4.2 hours. They found four new satellite candidates of NGC 7331 with low surface brightness and a faint stellar filament with a surface brightness of $27^m/arcsec^2$ in the g-band, located in the western outskirts of NGC 7331.

Members of the TBG group (Tief Belichtete Galaxien) in the FG Astrofotografie of the german Vereinigung der Sternfreunde e.V. (VdS) have achieved deep images of NGC 7331 and its surrounding. The used telescopes have apertures of 500 mm (7,5 h), 360 mm (4.0 h) and 110 mm (26 h), in brackets the exposure times in hours, respectively. On these images all four objects NGC 7331 A, B, C and D from [11] are visible and also a new object with low surface brightness, marked by "E" in our image (Fig. 1). This likely dwarf companion of NGC 7331 with the coordinates RA = $22^h37^m52.2^s$, Dec = +34°33'28" is found in a projected separation of 13.2' or 56 kpc. The angular diameter of the dwarf is 0.36' or 1.5 kpc. Based on the photometric magnitude measurements of A, B, C and D from [11] we have determined the magnitude of object E to be B = $19.5^m \pm 0.3^m$. The absolute magnitude $M_B = -11.6^m$ of this new dwarf galaxy and the linear diameter of 1.5 kpc seem to be typical for spheroidal companions of other nearby spiral galaxies.

In a projected separation of 12' or 52 kpc north of NGC 7331 there is the more luminous galaxy MAPS-PP O-778-973630 with the coordinates RA = $22^h37^m12,4^s$ and Dec = +34°37'13". In Fig. 1, we simply designate it as ("M"). Its visual magnitude is $16.15^m$ and the angular diameter 0.56' x 0.38'. The diffuse morphololy of this nucleated dwarf allows us to associate a type dEn. The mosaic image of this object by H. Strauß, P. Remmel und M. Blauensteiner with exposure times of 7.5 h, 4.0 h and 26 h is presented in fig. 2. In these images the faint periphery of dwarf M is clearly elongated in direction to NGC 7331 with a total projected length of about 17 kpc. Obviously this structure has formed by the influence of NGC 7331´s tidal forces. Simultaneously we could not find any value of radial velocity or distance for this dwarf in the actual literature. In order to prove its physical connection to NGC 7331 we took a spectrum with the 6-m telescope of BTA SAO RAN on November 6$^{th}$, 2016. For this observation the grating VPHG1200G was used, the exposure time was 2 x 20 minutes. In the spectrum of galaxy M the Balmer absorption lines $H_\beta$, $H_\gamma$ and the G-band are visible (Fig.3). The determined radial velocity of $V_h$ = 763 ± 30 km/s or $V_{LG}$ = 1066 km/s differs only by -53 ± 30 km/s from that one of NGC 7331. So it is justified to assume this dwarf to be a real satellite of NGC 7331.

The existence of the tidal influenced elongated morphology of galaxy M was already mentioned by Paudel & Ree [12]. These authors found two more analogous examples of partly destroyed satellite galaxies around NGC 2968 and NGC 4203 in a distance of 17.5 Mpc and 15.2 Mpc, comparable to the distance of NGC 7331. In all three cases the main galaxies are situated in small groups and the dwarfs, distorted by tidal forces, are situated in projected distances of ~25-50 kpc from the massive parental galaxy with a radial velocity difference less than ~200 km/s. All tidal tails are elongated in the direction of the main galaxy. Such an orientation is likely implied by nearly radial orbits of these satellites and by the absence of other massive galaxies in the neighbourhood, that could generate a tangential component of the dwarfs´ motion. It seems important to mention that the progress in observational technics aiming on extremely low surface brightness values of 28-30$^m$/arcsec$^2$ will lead to the detection of further dwarf galaxies influenced by massive neighbour galaxies. Toloba et al. [13] found characteristics of tidal induced destruction of the dwarf galaxy Scl-MM-Dw2 in the neighbourhood of the massive spiral NGC 253. Similar chracteristics of destruction were found in the case of M 31´s dwarf companion NGC 147 [14]. The tidal streams of both dwarfs are orientated along a line that connects them to their main galaxy.

As known, only a small fraction of dwarf satellites with projected separations of ~(50-60) kpc from the main galaxy show signs of tidal destructions. This fact is obviously induced by the character of the orbital motion of the satellites, as well as by the structure of the groups as a whole, influencing the excentricity of the orbits. In this sense the search and analysis of tidal stirred dwarf galaxies can be an important tool for investigations of the kinematics of small galaxy groups.

IDK thanks for the support from the Russian Science Foundation, grant 14-02-00965.

[1] Group "Tief Belichtete Galaxien" of Vereinigung der Sternfreunde e.V., Fachgruppe Astrofotografie, PO Box 1169, D-64629, Heppenheim, Germany

[2] Special Astrophysical Observatory, Nizhni Arkhyz, KchR, 369167, Russia

**Figures**

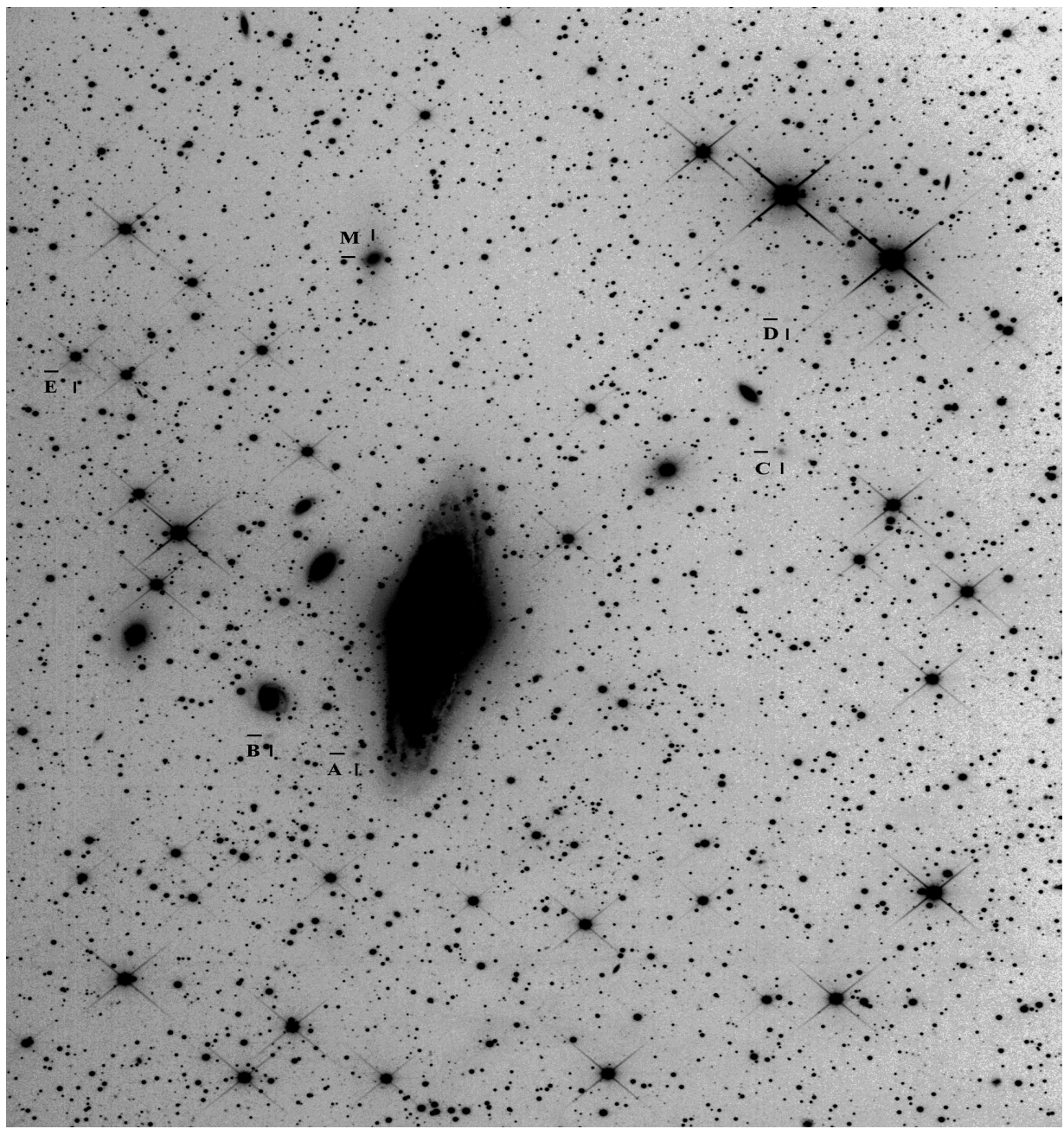

Fig. 1. The galaxy NGC 7331 and six of its supposed companions: NGC 7331 A, B, C, D, E and M. A 25.6' × 27.1' partial view of an image obtained by H. Strauß with an exposure time of 4 h. North is at the top, east is to the left.

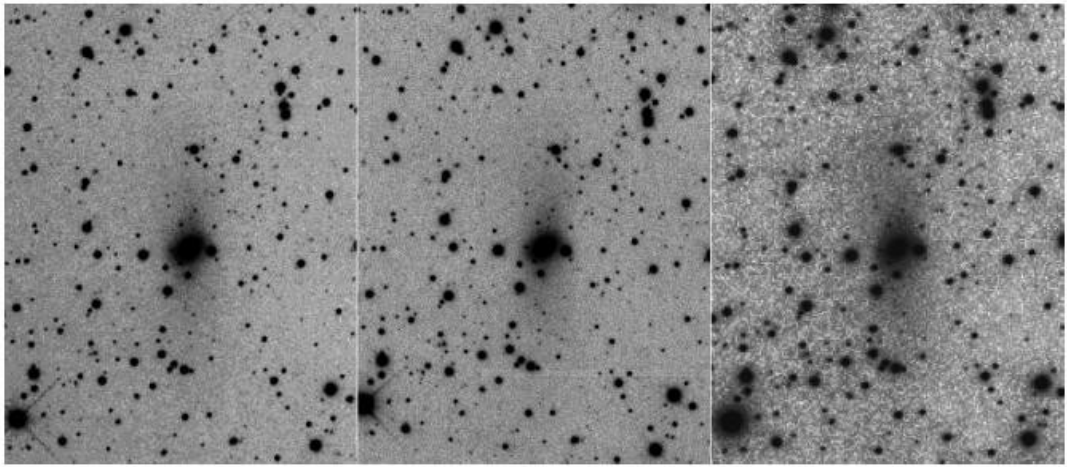

Fig. 2. Mosaic of three images of the disturbed dwarf galaxy N7331-dwM, obtained with different telescopes. North is at the top, east is to the left.

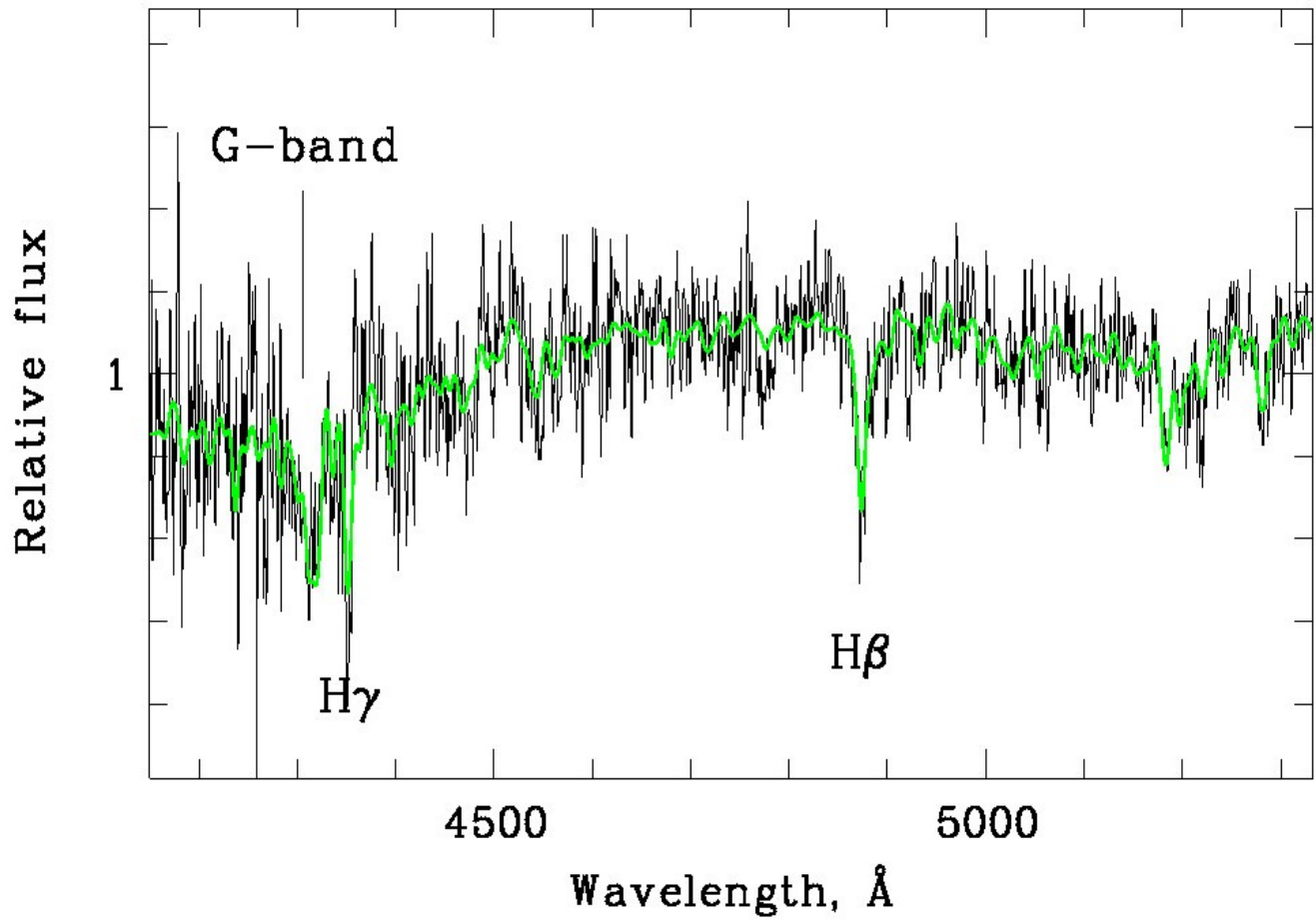

Fig. 3. Spectrum of the dwarf M obtained with the 6-m telescope.